# Topological energy transfer in an optomechanical system with exceptional points

H. Xu[1], D. Mason[1], L. Jiang[1] & J. G. E. Harris[1,2]

[1] *Department of Physics, Yale University, New Haven, Connecticut 06511, USA*

[2] *Department of Applied Physics, Yale University, New Haven, Connecticut 06511, USA*

**Topological operations have the merit of achieving certain goals without requiring accurate control over local operational details. To date, topological operations have been used to control geometric phases, and have been proposed as a means for controlling the state of certain systems within their degenerate subspaces[1–8]. More recently, it was predicted that topological operations can be extended to transfer energy between normal modes, provided that the system possesses a specific type of degeneracy known as an exceptional point (EP)[9–11]. Here we demonstrate the transfer of energy between two modes of a cryogenic optomechanical device by topological operations. We show that this transfer arises from the presence of an EP in the device's spectrum. We also show that this transfer is non-reciprocal[12–14]. These results open new directions in system control; they also open the possibility of exploring other dynamical effects related to EPs[15,16], as well as the behavior of thermal and quantum fluctuations in the vicinity of EPs.**

An externally imposed time-variation of the Hamiltonian $H$ of an otherwise isolated, conservative system provides a powerful means for controlling the system's evolution. If $H$ is varied sufficiently slowly, the adiabatic theorem states that a system prepared at some initial time $t_0$ in a non-degenerate normal mode of $H(t_0)$ will remain in the corresponding normal mode of the instantaneous $H(t)$ [17]. As a result, varying $H$ so as to execute a closed loop (in the space of parameters that define $H$) will return the system to its initial state, up to an overall phase. This phase was shown by Berry and others to include a contribution that is determined by a simple geometric property of the control loop[1–4]. The subsequent insight that the effect of such a topological operation (e.g., executing a closed control path) may be robust against small fluctuations in the control path has had a profound impact on many areas of theory and experiment[5–8, 18].

More recently, it was predicted[9–11] that topological operations may also be used to transfer energy between modes in systems that are subject to loss and/or gain. Specifically, energy transfer was predicted to occur for closed adiabatic control paths that enclose an exceptional point (EP, a form of degeneracy that can arise when the effective $H$ is non-Hermitian). It was also predicted[12–14] that such operations can be non-reciprocal in their dependence upon the system's initial conditions and on the control loop's sense of rotation about the EP. The possibility of using topological operations to control the energy distribution within a system while also inducing non-reciprocal behavior has attracted considerable attention[19–22]. Some features of EPs have been demonstrated in static measurements of spectra and eigenmodes[23, 24]. However, experiments to date have not realized topological or non-reciprocal dynamics that arises from encircling an EP.

Here we measure topological and nonreciprocal dynamics in an optomechanical system. We show that the system possesses an EP, and that external control parameters can be used to encircle the EP on time scales comparable to the lifetime of the system's excitations. We demonstrate that such topological operations can transfer energy, and that this energy transfer is non-reciprocal. When the control path is not adiabatic, the dynamics becomes more complicated; however we find quantitative agreement between experimental data and numerical simulations over the full range of measurements.

The system studied here consists of a silicon nitride membrane placed inside a high-finesse optical cavity[25]. The membrane's dimensions are 1 mm $\times$ 1 mm $\times$ 50 nm. Because it is almost perfectly square, the membrane's vibrational eigenmodes include nearly-degenerate pairs that are well-separated in frequency from all the other eigenmodes. We use this separation to focus on a nearly-degenerate pair with natural frequencies $\omega_1/2\pi = 788.024$ kHz and $\omega_2/2\pi = 788.487$ kHz. In the absence of laser light driving the optical cavity, these two modes are essentially uncoupled, and have very small damping rates ($\gamma_1/2\pi = 0.6$ Hz and $\gamma_2/2\pi = 1.4$ Hz).

When a laser excites the cavity, the resulting intracavity field $\alpha$ drives the membrane's vibrations via radiation pressure. At the same time, the membrane's vibrations detune the cavity and thereby modulate $\alpha$[25,26]. It is straightforward to integrate $\alpha(t)$ out of the full optomechanical equations of motion (see Supplemental Material), resulting in an effective equation of motion for just $c_1$ and $c_2$, the displacements of the membrane's modes:

$$i\dot{\vec{C}}(t) = H\vec{C}(t) \tag{1}$$

where $\vec{C}(t) = [c_1(t) \;\; c_2(t)]^T$. The effective Hamiltonian is:

$$H = \begin{pmatrix} \omega_1 - i\gamma_1/2 - ig_1^2\Sigma & -ig_1g_2\Sigma \\ -ig_1g_2\Sigma & \omega_2 - i\gamma_2/2 - ig_2^2\Sigma \end{pmatrix}, \tag{2}$$

where $g_{1,2}$ are the optomechanical coupling rates of the mechanical modes, and the complex mechanical susceptibility introduced by the intracavity field is $\Sigma = \frac{P}{\hbar\Omega_L}\frac{\kappa_{in}}{(\kappa/2)^2+\Delta^2}(\frac{1}{\kappa/2-i(\omega_0+\Delta)} - \frac{1}{\kappa/2+i(-\omega_0+\Delta)})$. Here $P$ and $\Omega_L$ are the power and frequency of the laser driving the cavity, $\Delta$ is the mean detuning between the laser and the cavity, $\omega_0 = (\omega_1 + \omega_2)/2$, and $\kappa$ and $\kappa_{\rm in}$ are the cavity's linewidth and input coupling rate. The phenomena studied here are classical; $\hbar$ appears in the expression for $\Sigma$ because $g_{1,2}$ are given in terms of the single-photon rate.

The system will possess an EP if $\Sigma$ can be made to equal $(\omega_1 - i\gamma_1/2 - \omega_2 + i\gamma_2/2)(-i(g_1^2 - g_2^2) \pm 2g_1g_2)/(g_1^2 + g_2^2)^2$. Achieving this typically requires control over both $\mathrm{Re}(\Sigma)$ and $\mathrm{Im}(\Sigma)$. For optomechanical devices in the resolved sideband regime ($\kappa < \omega_0$) this control is provided by $P$ and $\Delta$. In contrast, when $\kappa \gg \omega_0$, $P$ and $\Delta$ appear in $\Sigma$ in a linearly dependent fashion, and so only control $|\Sigma|$. The ability to access (and encircle) an EP using the detuning and power of a single laser is an important feature of the present system (and in contrast with the more complicated arrangement proposed in Ref. [27]), since these parameters can be controlled *in situ* with a high degree of precision, timing accuracy, and dynamic range.

A detailed description of the optomechanical device and the measurement setup is given in the Supplemental Material. The membrane and optical cavity are maintained at $T = 4.2$ K. The membrane's motion is monitored via a heterodyne measurement of a laser with constant power and detuning. Control over the optomechanical system is provided by a separate laser, whose detuning $\Delta$ and power $P$ are set by an acousto-optic modulator.

To establish the presence of an EP in this system, we measured the membrane's mechanical spectrum as a function of $\Delta$ and $P$. These spectra were acquired by driving the membrane and monitoring its response via the heterodyne signal. As described in the Supplemental Material, each spectrum was fit to determine the two resonance frequencies $\omega_{\mathrm{a,b}}(\Delta, P)$ and damping rates $\gamma_{\mathrm{a,b}}(\Delta, P)$. (The subscripts a,b (1,2) refer to the membrane's normal modes in the presence (absence) of an optical field.)

The results of these fits are summarized in Fig. 1, which shows the complex eigenvalues $\xi_{\mathrm{a,b}} = \omega_{\mathrm{a,b}} - i\gamma_{\mathrm{a,b}}/2$ as $\Delta$ and $P$ are varied. When $P \leq 155\ \mu$W, $\xi_a$ and $\xi_b$ each trace out a closed trajectory, completing a loop as $\Delta$ is varied from $\ll -\omega_0$ to $\gg -\omega_0$. In contrast, when $P \geq 265\ \mu$W, $\xi_a$ and $\xi_b$ both follow open trajectories, swapping their values as $\Delta$ is varied over the same range. This sharp transition in the topology of $\xi_{a,b}(\Delta)$ is characterstic of an EP[9]. The solid lines in Fig. 1 are a global fit to the complex eigenvalues of $H$, which gives best-fit values of $\omega_{1,2}$ and $\gamma_{1,2}$ as stated above, as well as $g_1/2\pi = 1.03$ Hz, $g_2/2\pi = 1.14$ Hz, $\kappa_{\mathrm{in}}/2\pi = 70$ kHz, and $\kappa/2\pi = 177$ kHz. These values imply an EP at $\Delta_{\mathrm{EP}}/2\pi = -792.5$ kHz, $P_{\mathrm{EP}} = 223\ \mu$W (or, equivalently $\omega_{\mathrm{EP}}/2\pi = 788.2$ kHz and $\gamma_{\mathrm{EP}}/2\pi = 460$ Hz, indicated as a black "$\times$" in Fig. 1).

Figure 2 (a) and (b) show measurements of $\mathrm{Re}(\xi_{a,b})$ and $-2\,\mathrm{Im}(\xi_{a,b})$ over a narrow range of $\Delta$ and $P$ centered on $\Delta_{\mathrm{EP}}$ and $P_{\mathrm{EP}}$. These measurements show the charateristic features of an EP:

$\xi_a$ and $\xi_b$ coalesce at a single value of the control parameters, and in the vicinity of this point they exhibit the same structure as the Riemann sheets of the complex square-root function $z^{1/2}$. For comparison, Fig. 2 (c) and (d) show the eigenvalues of $H$ (see Eq. 1), calculated using the best-fit values determined in Fig. 1.

The surfaces shown in Fig. 2 (a) and (b) are such that if $\Delta$ and $P$ were varied to execute a single closed loop, the resulting smooth evolution on the eigenvalue manifold would return to its starting point only if the loop did not enclose the EP. In contrast, a loop enclosing the EP would result in a trajectory starting on one sheet but ending on the other.

To observe this effect, we performed a series of measurements in which $\Delta$ and $P$ were initially set to $\Delta_{\max}$ and $P_{\min}$, and one of the membrane's modes ($c_{\mathrm{a}}$) was excited using a piezo-electric element. Once the system reached its steady state, the piezo drive was switched off, and $\Delta$ and $P$ were varied to sweep out a closed rectangular loop. The loop was defined by the points $\{\Delta_{\max}, P_{\min}\}, \{\Delta_{\max}, P_{\max}\}, \{\Delta_{\min}, P_{\max}\}, \{\Delta_{\min}, P_{\min}\}$, returning to $\{\Delta_{\max}, P_{\min}\}$ after a duration $\tau = 16$ ms. This value of $\tau$ was chosen to satisfy the requirement of conventional adiabaticity, namely $\tau \gg 1/|\xi_{\mathrm{a}} - \xi_{\mathrm{b}}|$ (except for the control loops that pass close to the exceptional point). Further below, we describe the effect of varying $\tau$.

The heterodyne signal was recorded before, during, and after the control loop. This signal was demodulated at frequencies $\omega_{\mathrm{a}}(\Delta_{\max}, P_{\min})$ and $\omega_{\mathrm{b}}(\Delta_{\max}, P_{\min})$, with typical results shown in Fig. 3 (a) and (b). Before and after the control loop (i.e., for $t < 0$ and $t > \tau$), this record corresponds to the amplitudes of the normal modes' motion $|c_{\mathrm{a}}(t)|$ (red data) and $|c_{\mathrm{b}}(t)|$ (blue data). During the control loop ($0 \leq t \leq \tau$) this correspondence does not hold, as the membrane's eigenfrequencies undergo rapid variations; data from this region does not play any role in our analysis. As shown in Fig. 3 (a) and (b), $c_{\mathrm{a}}$ is initially excited to $\sim 4 \times 10^{-12}$ m. There is also a small excitation of $c_{\mathrm{b}}$ (owing to the non-zero overlap of the mechanical resonances); however this unintentional excitation accounts for $\lesssim 1\%$ of the total energy, and does not qualitatively impact the results presented here.

Comparing $|c_{\mathrm{a,b}}(0)|$ with $|c_{\mathrm{a,b}}(\tau)|$ in Fig. 3 (a) and (b), it is clear that energy is lost from the system during the control loop. This reflects the fact that the damping here is always positive. To distinguish this overall energy loss from effects related to the topological operation, we focus on the relative energy of the two modes before and after the loop.

The data in Fig. 3 (a) was taken for a control loop not enclosing the EP ($\Delta_{\mathrm{max}} =$ -1,440 kHz, $P_{\mathrm{max}} = 750\ \mu$W; for all data, $\Delta_{\mathrm{min}} =$ -1,890 kHz, $P_{\mathrm{min}} = 2\ \mu$W). As a result, the nearly-adiabatic transit around the control loop results in the system returning to the same state at the end of the control loop. This can be seen qualitatively in Fig. 3 (a) by noting that $\approx 99\%$ of the energy is in $c_{\mathrm{a}}$ both immediately before and immediately after the control loop.

In contrast, Fig. 3 (b) shows a measurement in which the control loop does enclose the EP ($\Delta_{\mathrm{max}} = -300$ kHz, $P_{\mathrm{max}} = 750\ \mu$W). The impact on the dynamics is readily visible: before the loop, $> 99\%$ of the energy is in $c_{\mathrm{a}}$, while after the loop, $> 99\%$ of the (remaining) energy is in $c_{\mathrm{b}}$.

To quantify the transfer of energy from one mode to another, we define the efficiency $E =$

$|c_{\rm b}(\tau)|^2/(|c_{\rm a}(\tau)|^2 + |c_{\rm b}(\tau)|^2)$ (this definition makes use of the fact that prior to the loop, nearly all the energy is in $c_{\rm a}$). The values of $|c_{\rm a,b}(\tau)|$ are determined by fitting a decaying exponential to $|c_{\rm a,b}(t)|$ for $t > \tau + 20$ ms, and extrapolating these fits to $t = \tau$.

Figure 3 (c) shows $E(\Delta_{\rm max})$ for fixed $P_{\rm max} = 750\ \mu$W, while Fig. 3 (d) shows $E(P_{\rm max})$ for fixed $\Delta_{\rm max} = -290$ kHz. The limiting behavior in both cases (i.e., for large or small $P_{\rm max}$ and $\Delta_{\rm max}$) agrees with the prediction that adiabatic paths enclosing the EP will result in energy transfer, while adiabatic paths not enclosing the EP will not. The solid lines in Figs. 3 (c) and (d) are the results of numerically integrating Equations 1 and 2, and are not fits; rather, they use the same $P(t)$ and $\Delta(t)$ employed in the measurements, and the values of $g_{1,2}$, $\omega_{1,2}$, $\gamma_{1,2}$, $\kappa_{\rm in}$, and $\kappa$ determined from the data in Fig. 1. These simulations show good agreement with the measurements whether or not the loop encloses the EP, and whether or not the loop satisfies adiabiticity.

The measurements shown in Fig. 3 were all made by applying the initial drive to the "a" mode and then executing a control loop in the counter-clockwise (CCW) sense. In this case, the adiabatic trajectories enclosing the EP correspond to the less-damped eigenmode (red regions of the surfaces in Fig. 2) for the majority of the loop. In contrast, executing the same loop in the clockwise (CW) sense would result in an adiabatic trajectory corresponding primarily to the more-damped eigenmode (blue regions in Fig. 2). As described in Refs. [12–14, 28], adiabatic behavior is expected while the system is in the less-damped eigenmode; however, when the system is in the more-damped mode, competition between the non-adiabatic transfer (which is exponentially small in $\tau$) and the impact of differential loss (which is exponentially large in $\tau$) leads to a breakdown of adiabaticity, causing the system to eventually relax to the less-damped mode. This process may also be understood as a consequence of the Stokes phenomenon of asymptotics[12].

This behavior is demonstrated in Fig. 4, which shows $E(\tau)$ when the EP is encircled in the CCW or CW sense, and with the initial excitation in the "a" mode (for which $E$ is as defined above) or the "b" mode (for which $E$ is as defined above, but with the subscripts reversed). The same loop was used in all four cases: $\Delta_{\text{min}} = -1,890$ kHz, $P_{\text{min}} = 2$ $\mu$W, $\Delta_{\text{max}} = -290$ kHz, $P_{\text{max}} = 750$ $\mu$W. In all four cases, executing the loop very quickly results in negligible energy transfer (i.e., $E \to 0$ as $\tau \to 0$), consistent with the conventional expectation for a sudden perturbation.

The adiabatic limit ($\tau \gg 1$ ms) is quite different. Efficient energy transfer is achieved ($E \to 1$) for an initial excitation in the "a" mode and a CCW loop (and for an initial excitation in the "b" mode and a CW loop), consistent with the discussion of Fig. 3, and with the fact that these conditions correspond to adiabatic paths almost entirely in the less-damped mode. In contrast, $E \to 0$ when $\tau \gg 1$ ms for an initial excitation in the "b" mode and a CCW loop (and for an initial excitation in the "a" mode and a CW loop).

The behavior described above may be summarized by describing an adiabatic control loop around an EP as a matrix that transforms the initial state $[c_1(0)\ c_2(0)]^T$ to the final state $[c_1(\tau)\ c_2(\tau)]^T$ with the form:

$$U_{\circlearrowleft,\circlearrowright}(\tau) = \begin{pmatrix} a_{\circlearrowleft,\circlearrowright}(\tau) & b_{\circlearrowleft,\circlearrowright}(\tau) \\ c_{\circlearrowleft,\circlearrowright}(\tau) & d_{\circlearrowleft,\circlearrowright}(\tau) \end{pmatrix}, \tag{3}$$

where $\circlearrowleft, \circlearrowright$ denote a CCW and CW loop respectively. Because $H$ is a symmetric matrix, it is straightforward to show that, if $U_{\circlearrowleft}(\tau)$ and $U_{\circlearrowright}(\tau)$ represent identical but time-reversed control loops, then $U_{\circlearrowleft} = U_{\circlearrowright}^{T}$. Along with this relationship, the four data sets in Fig. 4 demonstrate the nonreciprocity of these operations, i.e. that $b_{\circlearrowleft,\circlearrowright}(\tau) \neq c_{\circlearrowleft,\circlearrowright}(\tau)$ for $\tau \gg 1$ ms[29]. This inequality is also evident in direct measurements of $|b_{\circlearrowleft,\circlearrowright}(\tau)|$ and $|c_{\circlearrowleft,\circlearrowright}(\tau)|$, as shown in the Supplemental Material.

In conclusion, we have demonstrated a new form of topological operation that allows for non-reciprocal energy transfer between two eigenmodes of a mechanical system. This transfer exploits the presence of an exceptional point in the two modes' spectrum. We note that the square membrane used in this work also offers three-fold and four-fold near-degneracies, opening the possibility of studying dynamics in the vicinity of higher-order exceptional points[15,16]. We also note that the cryogenic optomechanical device used in this demonstration is subject to both thermal and quantum fluctuations[30]; it is an open question whether non-reciprocal topological effects will allow for new forms of control over these fluctuations.

1. Simon, B. Holonomy, the quantum adiabatic theorem, and Berry's phase. *Physical Review Letters* **51**, 2167–2170 (1983).

2. Berry, M. V. Quantal phase factors accompanying adiabatic changes. *Proceedings of the Royal Society of London A: Mathematical, Physical and Engineering Sciences* **392**, 45–57 (1984).

3. Berry, M. V. Classical adiabatic angles and quantal adiabatic phase. *Journal of Physics A: Mathematical and General* **18**, 15 (1985).

4. Hannay, J. H. Angle variable holonomy in adiabatic excursion of an integrable Hamiltonian. *Journal of Physics A: Mathematical and General* **18**, 221 (1985).

5. Arovas, D., Schrieffer, J. R. & Wilczek, F. Fractional Statistics and the Quantum Hall Effect. *Physical Review Letters* **53**, 722–723 (1984).

6. Tomita, A. & Chiao, R. Y. Observation of Berry's topological phase by use of an optical fiber. *Physical Review Letters* **57**, 937–940 (1986).

7. Kitaev, A. Y. Fault-tolerant quantum computation by anyons. *Annals of Physics* **303**, 2–30 (2003).

8. Nayak, C., Simon, S. H., Stern, A., Freedman, M. & Das Sarma, S. Non-Abelian anyons and topological quantum computation. *Reviews of Modern Physics* **80**, 1083–1159 (2008).

9. Heiss, W. D. Phases of wave functions and level repulsion. *The European Physical Journal D-Atomic, Molecular, Optical and Plasma Physics* **7**, 1–4 (1999).

10. Keck, F., Korsch, H. J. & Mossmann, S. Unfolding a diabolic point: a generalized crossing scenario. *Journal of Physics A: Mathematical and General* **36**, 2125 (2003).

11. Berry, M. V. Physics of nonhermitian degeneracies. *Czechoslovak Journal of Physics* **54**, 1039–1047 (2004).

12. Berry, M. V. & Uzdin, R. Slow non-Hermitian cycling: exact solutions and the Stokes phenomenon. *Journal of Physics A: Mathematical and Theoretical* **44**, 435303 (2011).

13. Uzdin, R., Mailybaev, A. & Moiseyev, N. On the observability and asymmetry of adiabatic state flips generated by exceptional points. *Journal of Physics A: Mathematical and Theoretical* **44**, 435302 (2011).

14. Milburn, T. J. *et al.* General description of quasiadiabatic dynamical phenomena near exceptional points. *Physical Review A* **92**, 052124 (2015).

15. Cartarius, H., Main, J. & Wunner, G. Exceptional points in the spectra of atoms in external fields. *Physical Review A* **79**, 053408 (2009).

16. Demange, G. & Graefe, E.-M. Signatures of three coalescing eigenfunctions. *Jornal of Physics A: Mathematical and Theoretical* **45**, 025303 (2012).

17. Arnold, V. I. *Mathematical methods of classical mechanics*, vol. 60 of *Graduate Texts in Mathematics* (Springer New York, New York, NY, 1989).

18. Ando, T., Nakanishi, T. & Saito, R. Berry's phase and absence of back scattering in carbon nanotubes. *Journal of the Physical Society of Japan* **67**, 2857 (1998).

19. Lefebvre, R., Atabek, O., Šindelka, M. & Moiseyev, N. Resonance coalescence in molecular photodissociation. *Physical Review Letters* **103**, 123003 (2009).

20. Hamamda, M., Pillet, P., Lignier, H. & Comparat, D. Ro-vibrational cooling of molecules and prospects. *Journal of Physics B: Atomic, Molecular and Optical Physics* **48**, 182001 (2015).

21. Kaprálová-Žďánská, P. R. & Moiseyev, N. Helium in chirped laser fields as a time-asymmetric atomic switch. *The Journal of Chemical Physics* **141**, 014307 (2014).

22. Kim, S. Braid operation of exceptional points. *Fortschritte der Physik* **61**, 155–161 (2013).

23. Philipp, M., Brentano, P. v., Pascovici, G. & Richter, A. Frequency and width crossing of two interacting resonances in a microwave cavity. *Physical Review E* **62**, 1922–1926 (2000).

24. Dembowski, C. *et al.* Experimental observation of the topological structure of exceptional points. *Physical Review Letters* **86**, 787–790 (2001).

25. Thompson, J. D. *et al.* Strong dispersive coupling of a high-finesse cavity to a micromechanical membrane. *Nature* **452**, 72–75 (2008).

26. Aspelmeyer, M., Kippenberg, T. J. & Marquardt, F. Cavity optomechanics. *Reviews of Modern Physics* **86**, 1391–1452 (2014).

27. Lü, X.-Y., Jing, H., Ma, J.-Y. & Wu, Y. PT-Symmetry-Breaking Chaos in Optomechanics. *Physical Review Letters* **114**, 253601 (2015).

28. Graefe, E.-M., Mailybaev, A. A. & Moiseyev, N. Breakdown of adiabatic transfer of light in waveguides in the presence of absorption. *Physical review A* **88**, 033842 (2013).

29. Jalas, D. *et al.* What is — and what is not — an optical isolator. *Nature Photonics* **7**, 579–582 (2013).

30. Underwood, M. *et al.* Measurement of the motional sidebands of a nanogram-scale oscillator in the quantum regime. *Physical Review A* **92**, 061801 (2015).

**Acknowledgements** We thank Liang Jiang, Donghun Lee, Alexey Shkarin, and Woody Underwood for helpful discussions. This work was supported by AFOSR Grant FA9550-15-1-0270.

Figure 1: The complex eigenvalues of the membrane's normal modes. The resonance frequency (horizontal axis) and damping rate (vertical axis) of the membrane's two mechanical modes as a function of the laser power $P$ and detuning $\Delta$. Data for one mode is shown as squares, data for the other mode is shown as circles. The statistical uncertainty in the measurements is smaller than the symbols. Colour indicates $P$, while the arrows indicate the variation of the eigenvalues as $\Delta$ is changed from -1,890 kHz to -290 kHz at fixed $P$. For the lower values of $P$, each eigenvalue follows a closed trajectory, beginning and ending at the same point. For the higher values of $P$, the eigenvalues follow open trajectories, each one ending at the other's starting point. The solid lines are the global fit described in the text. The location of the EP predicted by this fit is shown as a black $\times$.

Figure 2: The EP in the spectrum of mechanical modes. a,b, The resonance frequencies (a) and damping rates (b) of the membrane's two mechanical modes as a function of $P$ and $\Delta$. Each grid point corresponds to a measurement; grid lines and surface colouring are guides to the eye. Colouring is chosen so that red (blue) corresponds to the mode with lower (higher) damping. c,d, Plot of the theoretically calculated real (c) and imaginary (d) parts of the eigenvalues of the effective Hamiltonian matrix $H$ (Eq. 2). All of the parameters appearing in this calculation are taken from the fit in Fig. 1. Note that the viewing angle in (a),(c) differs from that in (b),(d).

Figure 3: Topological energy transfer. a,b, The amplitude of motion of the two mechanical modes as a function of time $t$. A drive is applied to the "a" mode for $t < 0$. At $t = 0$ the drive is turned off and the control loop described in the text is implemented. The control loop ends at $t = 16$ ms. For $t > 16$ ms the system relaxes to thermal equilibrium. The black lines are fits to a decaying exponential (due to the mechanical damping) with a constant offset (reflecting the mode's thermal motion). The black dot shows the extrapolation of this fit to $t = 16$ ms. The loop used in (a) does not enclose the EP, while the loop used in (b) does. c, The fraction of the (remaining) energy in the "b" mode after the control loop has been completed, as a function of the loop's maximum detuning $\Delta_{\max}$. The data in (a) and (b) correspond to the two points shown as solid circles. d, The corresponding measurement as a function of the loop's maximum power $P_{\max}$. In both (c) and (d), the statistical errors are comparable to or smaller than the symbols. The solid lines are numerical simulations of the dynamics which are completely constrained by the parameters from the fit in Fig. 1. The insets are schematic illustrations showing how the loop varies along the horizontal axis of each panel; the location of the EP is indicated by the black $\times$.

Figure 4: Non-reciprocal topological dynamics. a-b, The transfer efficiency $E$ as a function of the control loop's duration $\tau$. The loop shape is identical in all four plots and encloses the EP. The loop is counter-clockwise in (a), while clockwise in (b). Red colour represents that the "a" mode is initially driven, while blue colour represents that the "b" mode is initially driven. In all four cases, rapid circulation around the loop ($\tau \to 0$) results in vanishing energy transfer ($E \to 0$). For adiabatic circulation, the limiting behavior of $E$ depends upon the sense of circulation and which mode is initially excited. For counter-clockwise (clockwise) loop, the red (blue) plot corresponds to conventional adiabaticity ($E \to 1$ as $\tau$ increases), while the blue (red) plot shows the opposite behavior ($E \to 0$ as $\tau$ increases). As described in the text, this reflects the non-reciprocity of each topological operation (counter-clockwise or clockwise loop). In both (a) and (b), the solid lines are numerical simulations of the dynamics which are completely constrained by the parameters from the fit in Fig. 1.

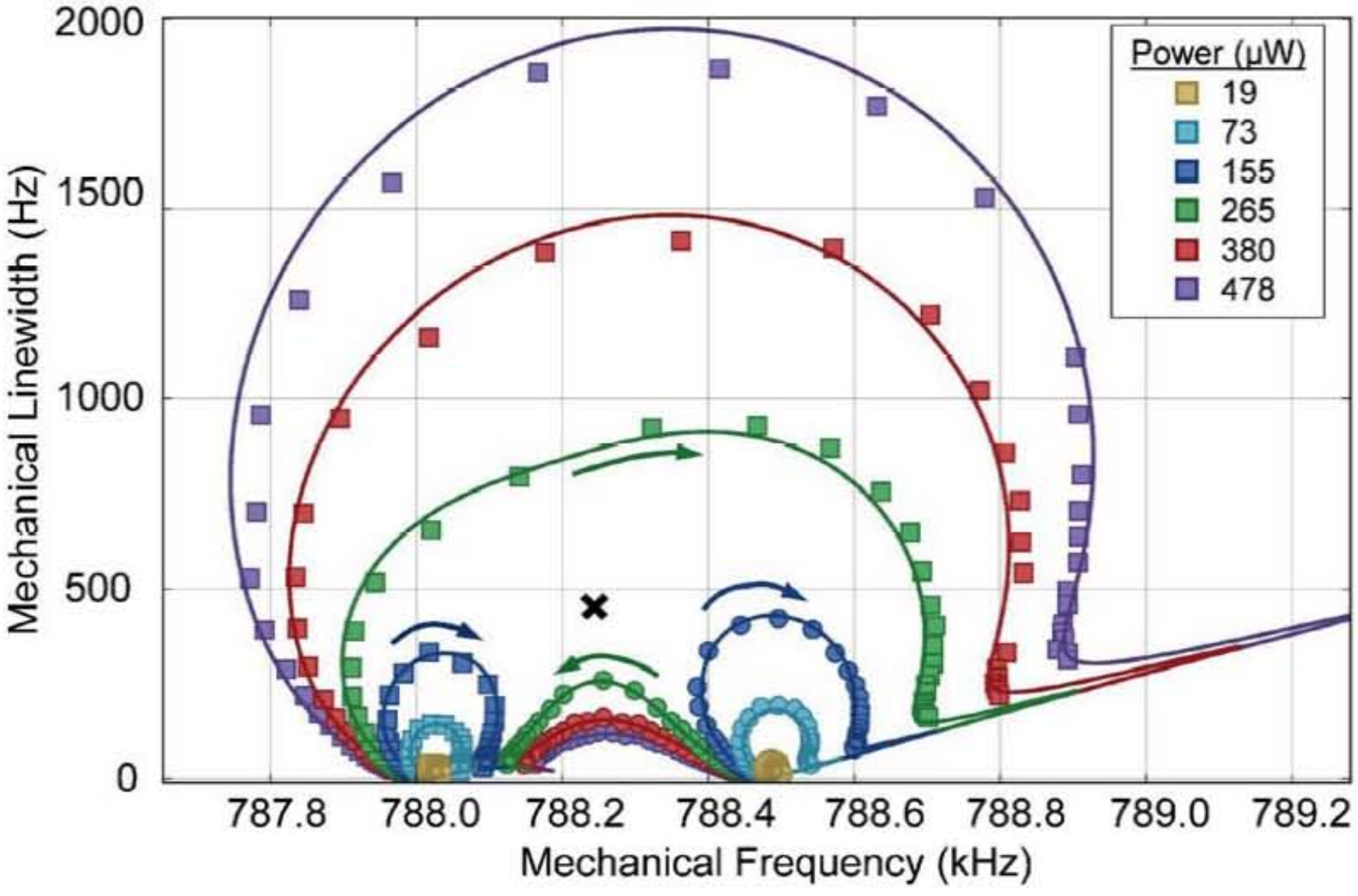

Fig. 1

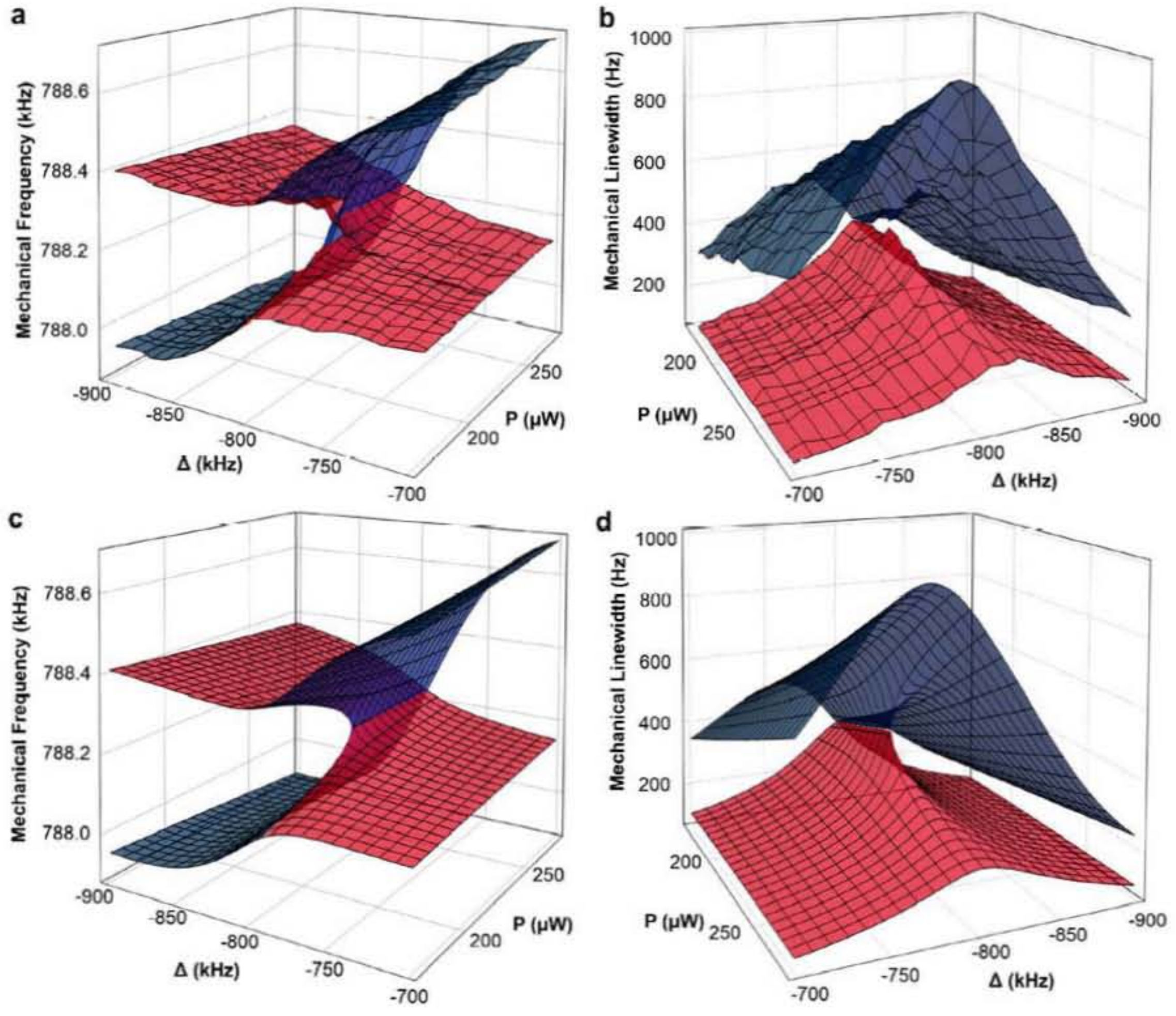

Fig. 2

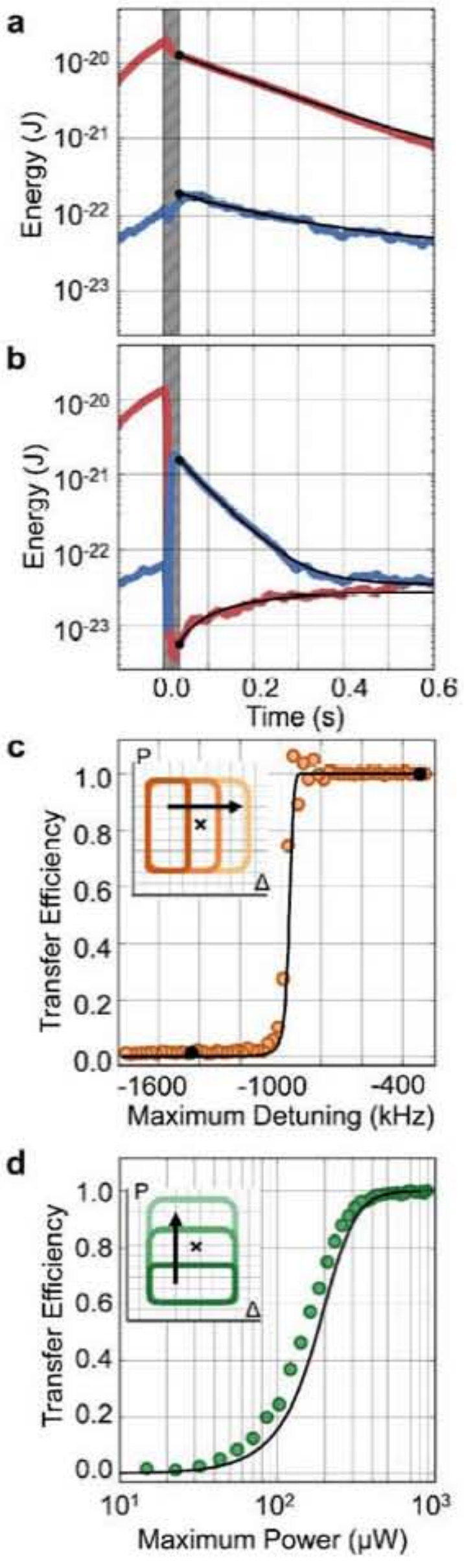

Fig. 3

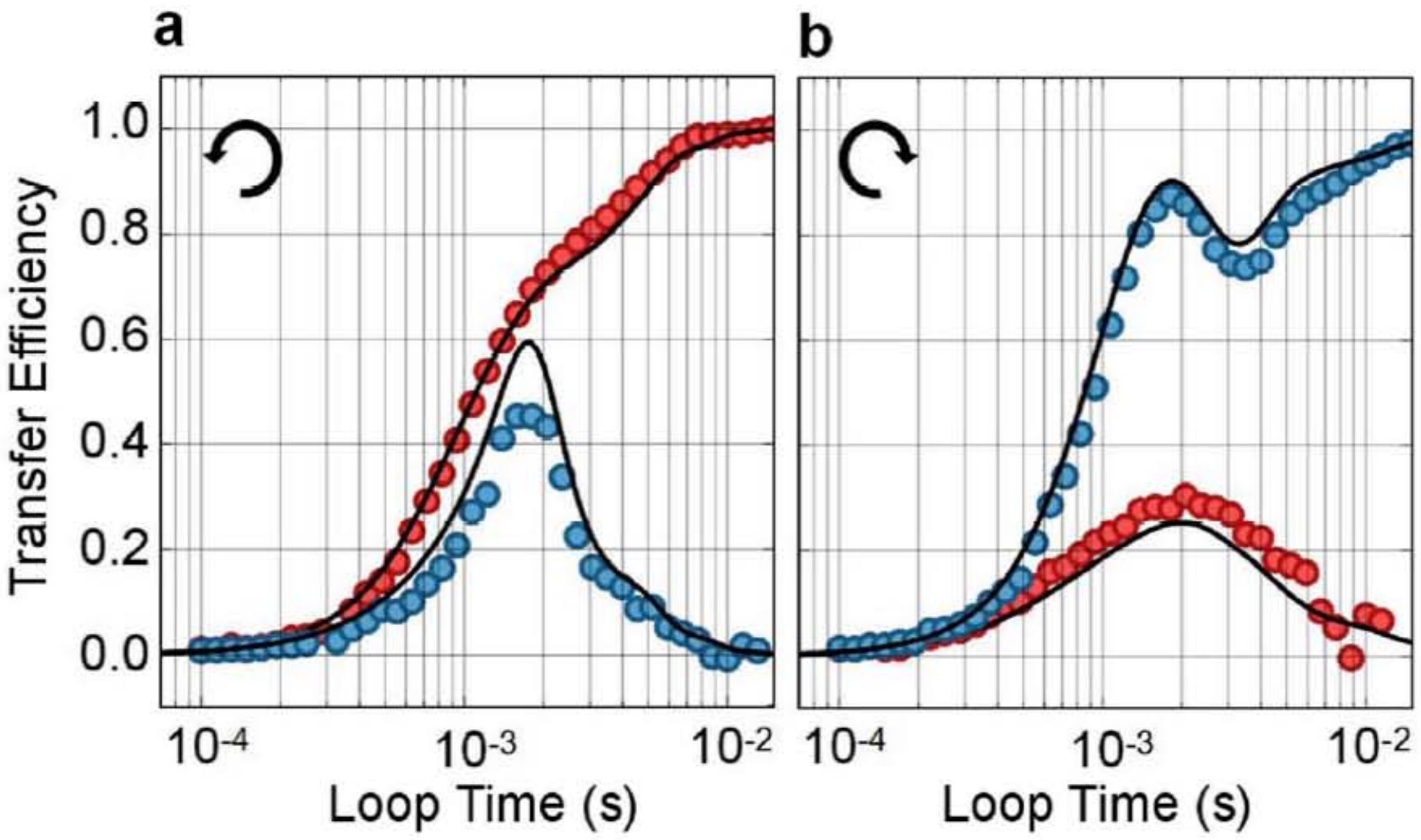

Fig. 4

# Supplementary Information

H. Xu,[1] D. Mason,[1] L. Jiang,[1] J. G. E. Harris[1,2]

[1]Department of Physics, Yale University, New Haven, Connecticut 06511, USA

[2]Department of Applied Physics, Yale University, New Haven, Connecticut 06511, USA

## 1 Measurement Setup

A schematic illustration of the experiment is shown in Fig. S1. The optomechanical device and much of the measurement setup are described in Ref. [1]. The membrane and optical cavity are mounted in a cryostat which is maintained at $T = 4.2$ K. The membrane's motion is monitored via a heterodyne measurement using a probe beam and a local oscillator (LO), both produced from a single laser ("ML" in Fig. S1 (a)). The probe beam frequency is shifted by an acousto-optical modulator (AOM1 in Fig. S1 (a)) driven at 80 MHz. Pound-Drever-Hall locking is used to keep the probe beam nearly resonant with one mode of the cavity; as a result its detuning $\Delta_{\rm p} \ll \kappa$, resulting in a negligible contribution to $\Sigma$. Likewise, the large detuning of the LO ($\Delta_{\rm LO} \approx 80$ MHz $\gg \kappa$) also results in a negligible contribution to $\Sigma$. Control over the optomechanical system is provided by a separate laser ("CL" in Fig. S1 (a)), whose detuning $\Delta$ and power $P$ are controlled by an additional acousto-optic modulator (AOM3 in Fig. S1 (a)). The frequencies of the various beams are illustrated in Fig. S1 (b). The cavity is approximately single-sided, and all measurements are performed in reflection. The reflected beams are incident on a single photodiode, and demodulation circuits are used to monitor multiple Fourier components of the heterodyne signal, each with a bandwidth equal to 50 Hz.

## 2 Optically-mediated mechanical coupling

Here, we consider a system consisting of two mechanical modes, each coupled linearly to a common optical mode. We will show that the optical field generates a tunable effective coupling between the mechanical modes, which can be exploited to produce an exceptional point as described in the main text. The model follows closely the one presented in [2].

In a standard optomechanical system, one considers an optical cavity mode whose frequency is linearly coupled to the position of a mechanical oscillator. An input-output approach to this system yields a pair of coupled differential equations for the two modes, which can be easily treated in the Fourier domain to understand the optical modification of the mechanical susceptibility. Here, we consider the simple extension of this model in which there are two mechanical modes, each coupled to the same optical mode. This yields the following system of equations for the mechanical/optical modes:

$$\dot{a} = -\left(\frac{\kappa}{2} + i\omega_c\right) a - ig_1 a z_1 - ig_2 a z_2 + \sqrt{\kappa_{in}} a_{in} \tag{1}$$

$$\dot{c}_1 = -\left(\frac{\gamma_1}{2} + i\omega_1\right) c_1 - ig_1 a^* a + \sqrt{\gamma_1}\eta_1 \tag{2}$$

$$\dot{c}_2 = -\left(\frac{\gamma_2}{2} + i\omega_2\right) c_2 - ig_2 a^* a + \sqrt{\gamma_2}\eta_2 \tag{3}$$

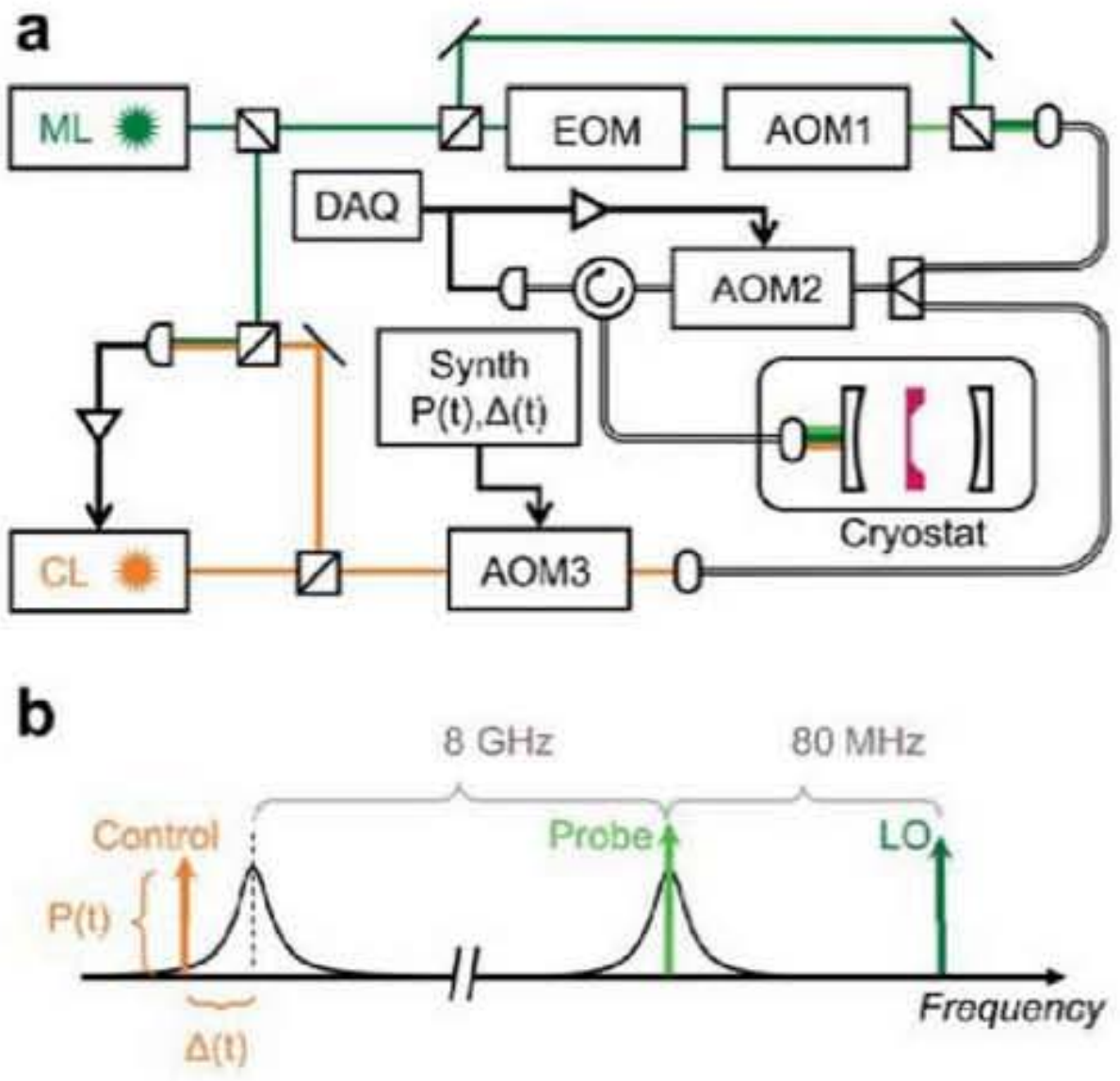

Figure S1: Experimental schematics. a, Illustration of the optical and electronic components. The measurement laser (ML) is split into a local oscillator (LO) and a probe beam using an acousto-optic modulator (AOM1). The probe beam is locked to the cavity using a Pound-Drever-Hall (PDH) scheme and modulation produced by an electro-optic modulator (EOM). The control laser (CL) is locked to the ML with a frequency offset approximately equal to twice the cavity's free spectral range. The control parameters used to access the EP are the CL's power $P$ and detuning $\Delta$. $P$ and $\Delta$ are set by the amplitude and frequency of an arbitrary waveform generator (AWG), which drives AOM3. The PDH error signal is used to control the frequency of AOM2, ensuring that all beams track fluctuations of the cavity. Light is delivered to (and collected from) the cryostat via an optical circulator. Coloured lines, hollow lines, and thick black lines show free-space laser beams, optical fibers, and electrical circuits respectively. Triangles, ovals, and semicircles show electronics, fiber couplers, and photodiodes respectively. The silicon nitride membrane is shown in purple. b, Illustration of the optical frequency domain. Lasers are coloured arrows and cavity modes are black curves.

where $a$ is the optical mode amplitude with resonant frequency $\omega_c$, total dissipation rate $\kappa$, and input coupling rate $\kappa_{in}$. The $i^{\text{th}}$ mechanical mode is described by position $z_i = c_i + c_i^*$ where $c_i$ is the complex mode amplitude. Each mechanical mode has resonant frequency $\omega_i$, dissipation rate $\gamma_i$, and is coupled to the optical mode with a single-photon coupling rate $g_i$. The optical and mechanical modes are driven by input fields $a_{in}$ and $\eta_i$, respectively.

We now suppose that the cavity is driven by a beam with power $P$ and frequency $\Omega_L$, detuned from the cavity resonance by $\Delta = \Omega_L - \omega_c$. Doing so, we can express the optical field as fluctuations $d(t)$ around a mean intracavity field given by $\bar{a} = \frac{\sqrt{\kappa_{in}}}{\frac{\kappa}{2} - i\Delta} a_{in}$ where $a_{in} = \sqrt{\frac{P}{\hbar\Omega_L}}$. Making these substitutions in the original system of equations yields the linearized equations of motion:

$$\dot{d} = -\left(\frac{\kappa}{2} - i\Delta\right) d - i\alpha_1 z_1 - i\alpha_2 z_2 \tag{4}$$

$$\dot{c}_1 = -\left(\frac{\gamma_1}{2} + i\omega_1\right) c_1 - i\left(\alpha_1^* d + \alpha_1 d^*\right) + \sqrt{\gamma_1}\eta_2 \tag{5}$$

$$\dot{c}_2 = -\left(\frac{\gamma_2}{2} + i\omega_2\right) c_2 - i\left(\alpha_2^* d + \alpha_2 d^*\right) + \sqrt{\gamma_2}\eta_2 \tag{6}$$

where we have defined $\alpha_i = \bar{a} g_i$. Moving to the Fourier domain, and defining the cavity susceptibility

$\chi_c[\omega] = (\frac{\kappa}{2} - i(\omega + \Delta))^{-1}$, we can solve for $d[\omega]$ and $d^*[\omega]$ and substitute these into the equations for $c_{1,2}[\omega]$ to find a reduced system of two equations describing the mechanical modes:

$$\left(\frac{\gamma_1}{2} - i(\omega - \omega_1)\right) c_1[\omega] = |\alpha_1|^2 \left(\chi_c^*[-\omega] - \chi_c[\omega]\right) c_1[\omega] + \alpha_1^* \alpha_2 \left(\chi_c^*[-\omega] - \chi_c[\omega]\right) c_2[\omega] \quad (7)$$

$$\left(\frac{\gamma_2}{2} - i(\omega - \omega_2)\right) c_2[\omega] = |\alpha_2|^2 \left(\chi_c^*[-\omega] - \chi_c[\omega]\right) c_2[\omega] + \alpha_1^* \alpha_2 \left(\chi_c^*[-\omega] - \chi_c[\omega]\right) c_1[\omega] \quad (8)$$

Note that we have dropped counter-rotating $c_1^*$ and $c_2^*$ terms. We have also dropped the mechanical drive terms $\eta_{1,2}$. These are not neccessary for our model, as we will simply drive the system to a particular initial state, turn off the drive, and focus on the evolution of the system without any mechanical drive applied.

In the traditional optomechanical system, one defines the optomechanical self-energy as $\mathbf{\Sigma}[\omega] = i|\alpha|^2(\chi_c^*[-\omega] - \chi_c[\omega])$. We see that in this 2-mode system, we can extend this concept to a self-energy matrix

$$\mathbf{\Sigma} = \begin{pmatrix} i|\alpha_1\alpha_1| & i|\alpha_1\alpha_2| \\ i|\alpha_2\alpha_1| & i|\alpha_2\alpha_2| \end{pmatrix} (\chi_c^*[-\omega] - \chi_c[\omega]) \quad (9)$$

Note that this definition of $\mathbf{\Sigma}$ differs slightly from the $\Sigma$ defined in the main text. Specifically, the dependence on $\alpha_1$ and $\alpha_2$ has been factored out, along with a factor of $-i$, leaving $\Sigma$ as a scalar quantity.

Writing our mechanical modes as a vector $\bar{c}[\omega] = \begin{pmatrix} c_1[\omega] \\ c_2[\omega] \end{pmatrix}$, we can write the following matrix equation:

$$-i\omega\bar{c}[\omega] = -\begin{pmatrix} \frac{\gamma_1}{2} + i\omega_1 & 0 \\ 0 & \frac{\gamma_2}{2} + i\omega_2 \end{pmatrix} \bar{c}[\omega] - i\mathbf{\Sigma}[\omega]\bar{c}[\omega] \quad (10)$$

Before we can move back to the time domain, we note that $\Sigma$ varies on the scale of $\kappa$, while the mechanical modes are only susceptible to drives within their linewidth, which is significantly smaller than $\kappa$, by assumption. Therefore, it is sufficient to consider $\mathbf{\Sigma}[\omega] \approx \Sigma[\omega_1] \approx \Sigma[\omega_2] \equiv \mathbf{\Sigma}$. (Note that the mechanical modes are also assumed to be nearly-degenerate). Now that $\mathbf{\Sigma}$ is not a function of $\omega$, we can easily move back to the time domain to find the equation from the main text:

$$i\dot{\bar{c}} = \boldsymbol{H}\bar{c} \quad (11)$$

where we define

$$\boldsymbol{H} = \begin{pmatrix} \omega_1 - i\frac{\gamma_1}{2} & 0 \\ 0 & \omega_2 - i\frac{\gamma_2}{2} \end{pmatrix} + \mathbf{\Sigma} \quad (12)$$

It is worth emphasizing that $\mathbf{\Sigma}$ is a complex quantity, which depends (via $\alpha_1$ and $\alpha_2$) on $P$ and $\Delta$. This is the tunability that allows us to access an exceptional point in the spectrum of these two mechanical modes.

# 3 Measuring the mechanical eigenvalue spectrum

In Figs. 2 and 3 in the main text, we show the complex eigenvalues (frequencies and decay rates) of the mechanical modes as a function of $P$ and $\Delta$. At each point $\{P, \Delta\}$, these eigenvalues were measured by optically driving the mechanical modes and measuring their driven response. We measure the mechanical sidebands using the heterodyne measurement laser, locked to the cavity resonance. We set a certain $P$ and $\Delta$ for the control laser, then apply amplitude modulation at a frequency near $\omega_1$ and $\omega_2$, thus creating an optical beat note which will drive the mechanical modes. This modulation frequency is swept over $\omega_1$ and $\omega_2$, and we use a lock-in amplifier to measure the complex response of the heterodyne signal to this drive.

Two examples of these measurements are shown below. Fig. S2 shows a sweep over the two modes when the control beam power is low, and there is minimal hybridization of the two modes. In Fig. S3, the control beam power is large and detuned near $-\omega_{1,2}$, such that the modes hybridize significantly, resulting in modes with degenerate frequencies but different linewidths. The relative phase of the driven response of the two modes is such that we see destructive interference in Fig. S3. By fitting the complex response to a sum of complex Lorentzians with an arbitrary phase offset, we extract $\omega_1, \omega_2, \gamma_1$, and $\gamma_2$. The solid lines in Figs. S2 and S3 are fits, from which we extract the eigenvalues plotted in Figs. 2 and 3 in the main text.

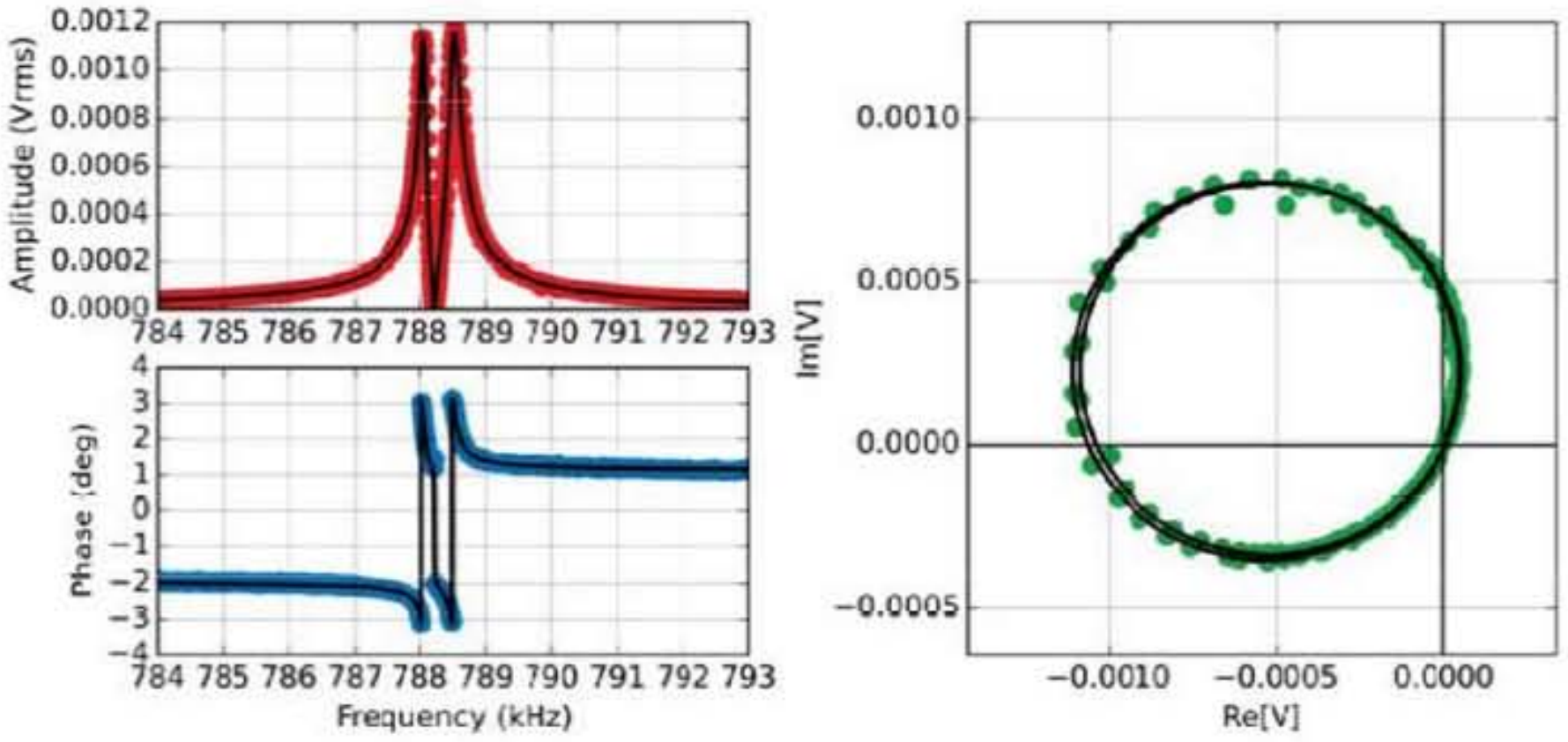

Figure S2: Low-Power Sweep ($\Delta = -780$ kHz, $P = 73$ $\mu$W). Lock-in signal as a function of drive frequency. Left panels: Amplitude (red) and Phase (blue) of the signal. Right panel: parametric plot of the in-phase and out-of-phase response.

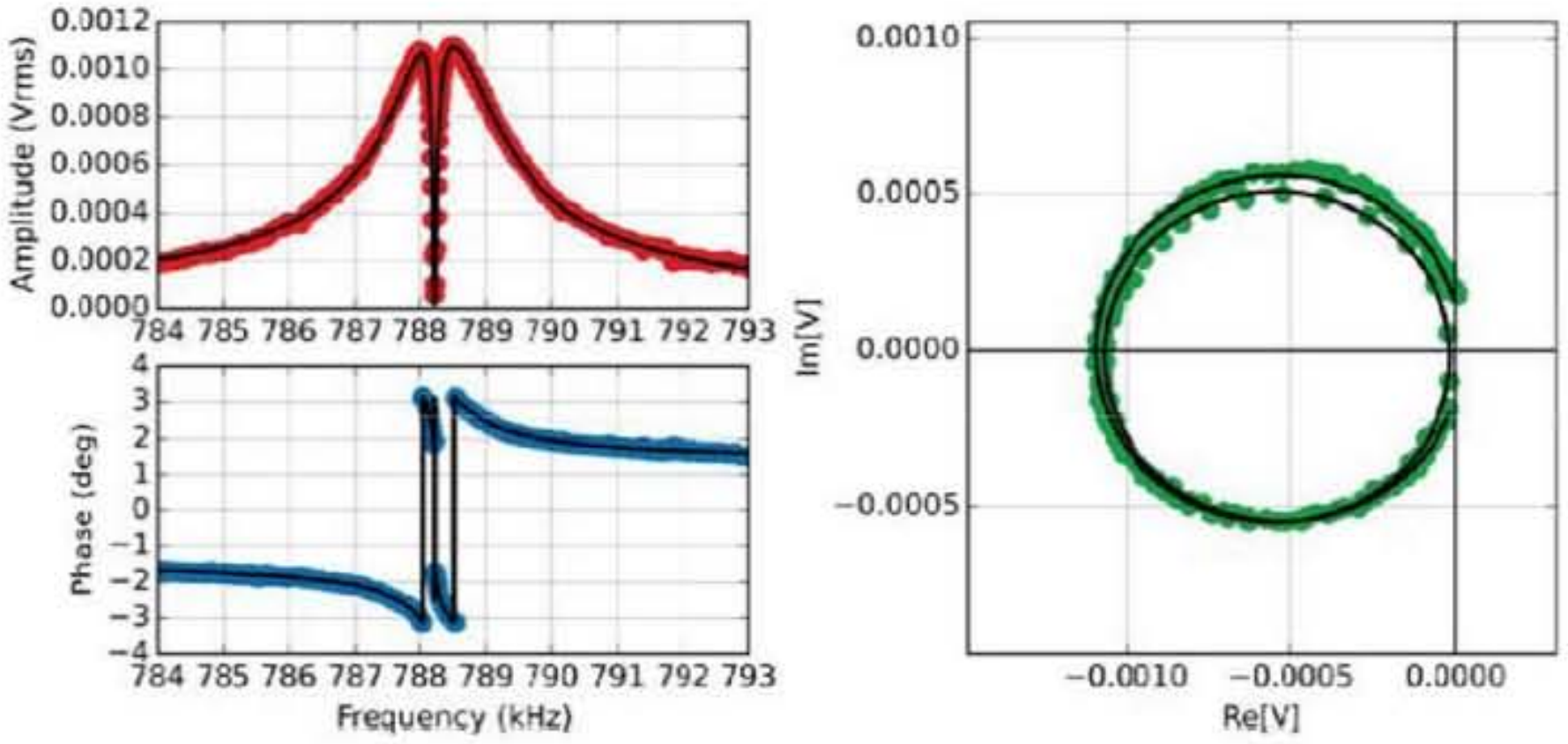

Figure S3: High-Power Sweep ($\Delta = -780$ kHz, $P = 380$ $\mu$W). Lock-in signal as a function of drive frequency. Left panels: Amplitude (red) and Phase (blue) of the signal. Right panel: parametric plot of the in-phase and out-of-phase response.

# 4 Measurement of Propagator Matrix Elements

In the main text, we describe energy transfer in terms of an efficiency defined as $\frac{|c_i(\tau)|^2}{|c_a(\tau)|^2+|c_b(\tau)|^2}$, where $i = \mathrm{a,b}$ depending on whether energy is being transferred to mode a or b. This parameter characterizes the relative energy transfer (independent of overall energy decay), but does not fully describe the effect of the control loop. In order to fully characterize the propagators $U_{\circlearrowleft}(\tau)$ and $U_{\circlearrowright}(\tau)$ (defined in the main text), we can instead examine the amplitude of motion in each mode before and after the control loop. So, for example, if we consider clockwise control loops in which the a mode is initially driven, then we can extract $|a_{\circlearrowright}(\tau)| = |\frac{c_a(\tau)}{c_a(0)}|$ and $|c_{\circlearrowright}(\tau)| = |\frac{c_b(\tau)}{c_a(0)}|$. Similarly, repeating this process with the initial excitation in the b mode gives $|b_{\circlearrowright}(\tau)|$ and $|d_{\circlearrowright}(\tau)|$. In Fig. S4, we plot these propagator matrix elements as a function of control loop duration, $\tau$. This is extracted from the same data as Fig. 4 in the main text.

For sufficiently large $\tau$, we see that $|b_{\circlearrowright,\circlearrowleft}(\tau)| \neq |c_{\circlearrowright,\circlearrowleft}(\tau)|$, which implies $b_{\circlearrowright,\circlearrowleft}(\tau) \neq c_{\circlearrowright,\circlearrowleft}(\tau)$, as stated in the main text. This allows the conclusion that $U_{\circlearrowright,\circlearrowleft}(\tau) \neq U^T_{\circlearrowright,\circlearrowleft}(\tau)$.

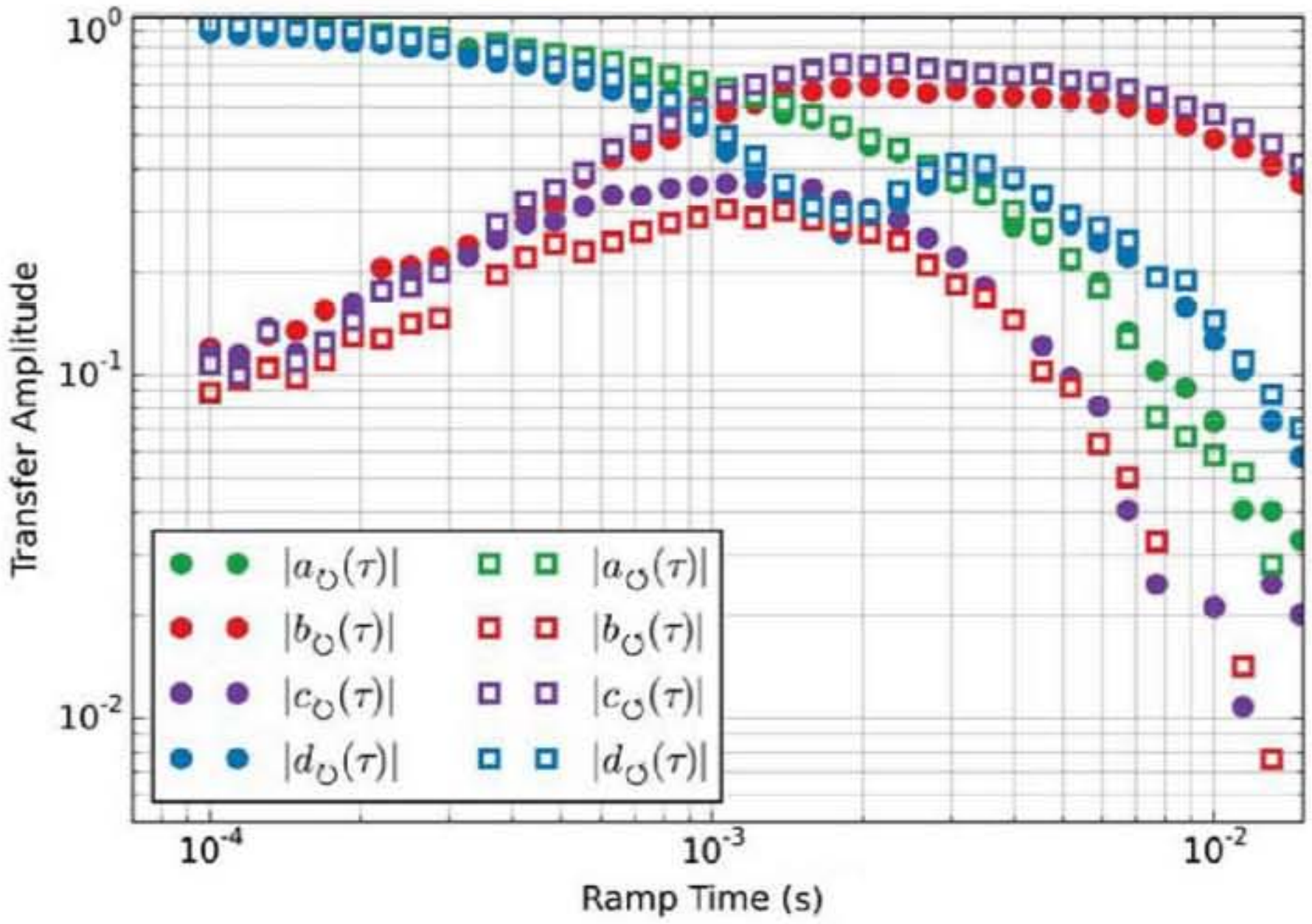

Figure S4: Magnitude of propagator matrix elements

# References

[1] M. Underwood, D. Mason, D. Lee, H. Xu, L. Jiang, A. B. Shkarin, K. Børkje, S. M. Girvin, and J. G. E. Harris, "Measurement of the motional sidebands of a nanogram-scale oscillator in the quantum regime," *Physical Review A - Atomic, Molecular, and Optical Physics*, vol. 92, no. 6, pp. 1–5, 2015.

[2] A. B. Shkarin, N. E. Flowers-Jacobs, S. W. Hoch, A. D. Kashkanova, C. Deutsch, J. Reichel, and J. G. E. Harris, "Optically mediated hybridization between two mechanical modes," *Physical Review Letters*, vol. 112, no. 1, pp. 1–5, 2014.